\documentclass[twocolumn,secnumarabic,amssymb,nobibnotes,aps,prl]{revtex4}

\usepackage{amsmath}
\usepackage{latexsym}
\usepackage{float}
\usepackage{amssymb}
\usepackage{graphicx}
\usepackage{textcomp}
\usepackage{hyperref}
\def\ba{\begin{eqnarray}}
\def\ea{\end{eqnarray}}
\def\be{\begin{equation}}
\def\ee{\end{equation}}

\def\bm{\begin{math}}
\def\me{\end{math}}

\newcommand{\dummy}

\begin{document}

\title{Domain Coarsening in 2-d Ising Model: Finite-Size Scaling for Conserved Dynamics}
%\vskip 0.5cm
\author{Suman Majumder and Subir K. Das$^{*}$}
\affiliation{Theoretical Sciences Unit, Jawaharlal Nehru Centre for Advanced Scientific Research, Jakkur P.O, Bangalore 560064, India}

\date{\today}

\begin{abstract}
We quantify the effect of system size in the kinetics of domain growth in Ising model with 50:50 composition in two spatial dimensions. Our estimate of the exponent, $\alpha=0.334\pm0.004$, for the power law growth of linear domain size, from Monte Carlo simulation using small systems of linear dimensions $L=16$, $32$, $64$, and $128$, is in excellent agreement with the prediction of Lifshitz-Slyozov (LS) theory, $\alpha=1/3$. We find that the LS exponent sets in very early and continues to be true until average size of domains reaches three quarters of equilibrium limit. 
\end{abstract}


\maketitle
 A homogeneous binary mixture, A+B, becomes unstable when quenched below the co-existence curve and moves towards the new equilibrium state via formation and growth of domains rich either in A- or B-particles  \cite{1}. This domain coarsening is a scaling phenomena, e.g., the two-point equal time correlation function, $C(r,t)$, which characterizes the domain morphology and growth, exhibits the scaling behavior \cite{2}
\begin{eqnarray}\label{1}
C(r,t) \equiv \tilde {C}(r/\ell(t)).
\end{eqnarray}
In Eq.(\ref{1}), $\tilde {C}(x)$ is a scaling function independent of the average domain length $\ell(t)$ which grows with time as
$\ell(t)\sim t^{\alpha}$. Associating the rate of domain growth with the gradient of chemical potential, one can write \cite{1}
\begin{eqnarray}\label{2}
 \frac{d\ell(t)}{dt} \sim \lvert  \overrightarrow{\nabla} \mu \lvert \sim \frac{\sigma}{\ell(t)^{2}},
\end{eqnarray}
where $\sigma$ is the A-B interfacial tension. Solving Eq.(\ref{2}) one gets $\alpha=1/3$, known as the Lifshitz-Slyozov (LS) \cite{3} growth law.
\par
While recent focus has been in more realistic systems \cite{4}, understanding of kinetics of phase separation even in simple spin-$\frac{1}{2}$ Ising model, where an up spin corresponds to A-particle and a down spin to B-particle, appears to be incomplete. For conserved order parameter, where composition of A- and B-species remains fixed during the entire evolution, dynamics in the Ising model is implemented via Kawasaki exchange mechanism \cite{5,6} where for a Monte Carlo (MC) move interchange of positions between a pair of nearest neighbor (nn) spins is tried. Earlier studies \cite{7} of phase ordering in conserved Ising model with 50:50 composition, most of which were based on MC simulation for very short period, reported estimates of $\alpha \in [0.17,0.25]$, deviating drastically from expected LS law. Even arguments in favor of logarithmic growth were proposed \cite{8}. 
\par
To resolve the controversy, Huse \cite{9} argued for the need to add an additional term $\propto$ $1/\ell(t)^{3}$, to Eq.(\ref{2}), accounting for an enhanced interface conductivity, which brings in a correction to the instantaneous exponent at finite $\ell(t)$ as
\begin{eqnarray}\label{3}
 \alpha_{i}=\frac{1}{3}[1-\frac{C_{1}}{\ell(t)}+O(\ell^{-2}(t))].
\end{eqnarray}
Indeed, consistency of growth exponent with the LS law was established for 50:50 binary mixture \cite{10} as well as multi-component mixture \cite{11}, in the limit $\ell(t \rightarrow \infty)$ $\rightarrow \infty$. Present work, however, shows that observation of LS value of the exponent only for large $\ell(t)$ is due to the mixing of a time dependent bare length in $\ell(t)$.
\par
Most of the studies, till date, stressed on using large systems, with the anticipation of a strong finite-size effect combined with the expectation that the LS law will be realized only in the large $\ell(t)$ limit, and, of course, to obtain better self-averaging. The objective of this paper is to study finite-size effect on domain coarsening in $2$-$d$ conserved Ising model and understand the behavior of $\alpha_{i}$ from appropriate scaling analysis \cite{6,12}. First effort in this direction was by Heermann, Yixue and Binder \cite{13} for extreme asymmetrical composition, however, was never followed up systematically. In this paper, we show via application of finite-size scaling method that the LS value of $\alpha$ sets in very early and effect of size is rather small so that using a system of size as small as $L^{2}=32^{2}$, one can confirm the LS growth law unambiguously. Such knowledge will be useful for computer simulation of more realistic systems by avoiding unnecessary large systems, thereby accessing large time scale. We also observe that systems do not provide self-averaging proportionate to their sizes\cite{14}.
\par
\begin{figure}[htb]
\centering
\includegraphics*[width=0.48\textwidth]{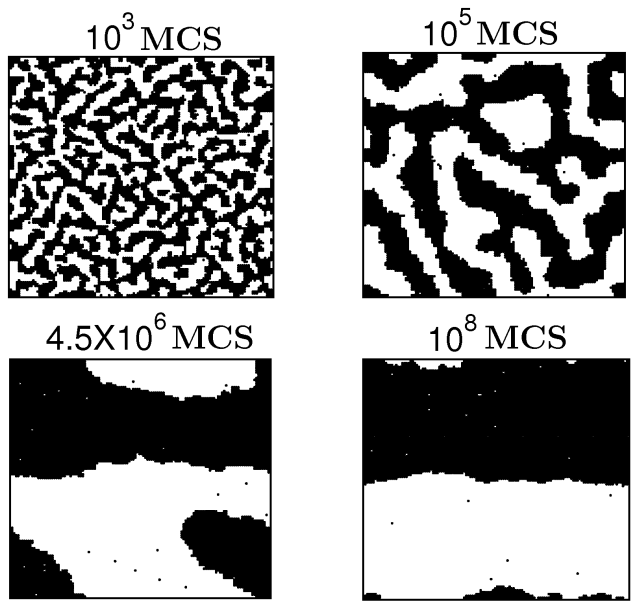}
\caption{\label{fig1} Evolution snapshots of domain coarsening in 2-d conserved Ising model following a quench from high temperature random state to $T=0.6T_{c}$, for a system of size $L^{2}=128^{2}$. The black dots mark the location of A-particles while the B-particles are unmarked.}
\end{figure}

In  Fig.\ref{fig1}, we present snapshots during the evolution of an Ising system, starting from a  50:50 random mixture of up and down spins, on a square lattice of linear size $L=128$, obtained via MC simulation at temperature $T=0.6T_{c}$, $T_{c}$ being the critical temperature. The times at which the shots were taken are mentioned on the figure in units of Monte Carlo steps (MCS) where each MCS consists of exchange trial over $L^{2}$ pair of spins. Periodic boundary conditions were applied in both $x$- and $y$- directions. While the last snapshot corresponds to a situation when A and B phases are completely separated, the one at $t=4.5\times10^{6}$ MCS represents the situation when finite-size effect began to enter, which will be clear from subsequent discussion. Note that all physical quantities were calculated from pure domain morphology after eliminating the thermal noise via a majority spin rule where a spin at a lattice site $j$ was replaced by the sign of majority of the spins sitting at $j$ and nn of $j$.
\par
\begin{figure}[htb]
\centering
\includegraphics*[width=0.48\textwidth]{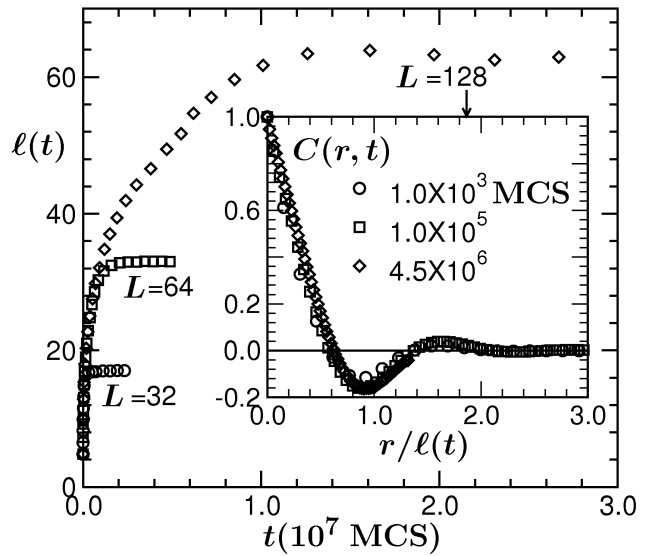}
\caption{\label{fig2} Plot of average domain size as a function of time for three different system sizes $L=32,64$, and $128$, as indicated. The inset shows a scaling plot of $C(r,t)$ vs $r/\ell(t)$ for $L=128$, for three different times, as mentioned.}
\end{figure}

Fig.\ref{fig2} shows the plot of $\ell(t)$ vs $t$ for $L=32,64$, and $128$, where $\ell(t)$ was calculated from the first moment of the domain-length distribution function \cite{11} $P(\ell_{k})$ with length $\ell_{k}$ being the separation between two successive interfaces in $x$- or $y$- directions. For $L=32$, the data were averaged over $2000$ independent initial configurations, for $L=64$ and $128$, averaging were done for $1000$ and $40$ initial configurations, respectively. The flat regions of the data sets correspond to the situation when the systems reached their final equilibrium states, thus domains cannot grow beyond this. This limiting value, for present method of calculation, comes out to be $\simeq$ $L/2=\ell_{\max}$. The last snapshot in Fig.\ref{fig1} represents such a situation. In the inset of Fig.\ref{fig2}, we demonstrate the scaling behavior of the correlation function, as embodied in Eq.(\ref{1}), for $L=128$. The data collapse upon dividing $r$ by $\ell(t)$ is good starting from as early as $t=10^{3}$ MCS till $t=4.5\times 10^{6}$ MCS when the finite-size effect begins. Apparently, as is clear from the plot of $\ell(t)$ vs $t$ as well as the scaling behavior of $C(r,t)$ for very extended time and length scale, the size effect is negligible \textit{almost} upto $\ell_{\max}$. However, to make quantitative statement about the extent of finite-size effect, appropriate scaling analysis is called for.
\par
In analogy with critical phenomena, one can construct the finite-size scaling ansatz \cite{13,15} by identifying $\ell(t)$ with the equilibrium correlation length $\xi$ and $1/t$ with temperature deviation from the critical point. At this stage we would like to quantify the growth of $\ell(t)$ as
\begin{eqnarray}\label{4}
\ell(t)=\ell(t_{0})+At^{\alpha},
\end{eqnarray}
where, to $\ell(t_{0})$, we assign the meaning of initial domain size when the system becomes unstable at time $t_{0}$ after the quench and the measurement of $t$ starts from $t_{0}$. Note that slightly poor data collapse in Fig.\ref{fig2} for the earliest time is primarily because $C(r,t)$ should have been plotted as a function of $r/[\ell(t)-\ell(t_{0})]$, not $r/\ell(t)$. But $C(r,t)$ contains information about $\ell(t)$ and subtracting the influence of $\ell(t_{0})$ is a rather challenging task. 
\par
Eq.(\ref{4}) is valid in absence of any finite-size effect. However, when $\ell(t)$ is comparable to $\ell_{\max}$, finite-size effect comes in and Eq.(\ref{4}) needs to be modified by accounting for the size effect as
\begin{eqnarray}\label{5}
 \ell(t)-\ell(t_{0})=y(x)t^{\alpha}.
\end{eqnarray}
In Eq.(\ref{5}), $y(x)$ is a scaling function, independent of the system size, which depends upon the scaling variable $x=(\ell_{\max}-\ell(t_{0}))/t^{\alpha}$. By closely observing Eqs. (\ref{4}) and (\ref{5}), one can write down the following behavior of $y(x)$ in the limiting situations:
\begin{eqnarray}\label{6}
 \lim_{~~~~~~~x \rightarrow 0(t \rightarrow \infty;\ell_{\max}<\infty)} y(x) = x,
\end{eqnarray}
\begin{eqnarray}\label{7}
 \lim_{~~~~~~~x \rightarrow \infty (\ell_{\max} \rightarrow \infty)} y(x) = A.
\end{eqnarray}

\par
\begin{figure}[htb]
\centering
\includegraphics*[width=0.48\textwidth]{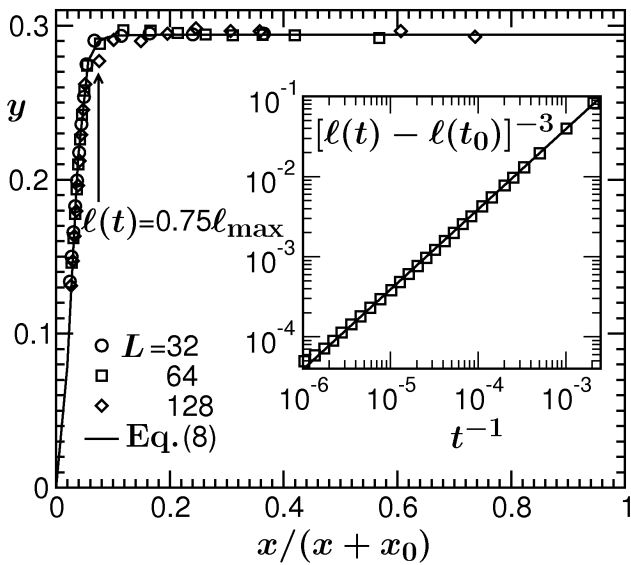}
\caption{\label{fig3} Finite-size scaling plot of $y(x)$, with $\ell(t_{0}=20)\simeq3.6$ and $\alpha=0.334$, as a function of $x/(x+x_{0})$; $x_{0}=5$. The continuous curve is a fit to Eq.(\ref{8}) with the best fit parameters mentioned in the text. The arrow roughly marks the appearance of finite-size effect. Inset: $[\ell(t)-\ell(t_{0})]^{-3}$ vs $t^{-1}$, for $L=64$, where the continuous line is a straight one with slope $39$.}
\end{figure}

 In Fig.\ref{fig3}, we plot $y(x)=[\ell(t)-\ell(t_{0})]/t^{\alpha}$ as a function of $x/(x+x_{0})$ with $x_{0}=5$, for which we have varied $\alpha$ and $\ell(t_{0})$ to get optimum collapse of data from different system sizes which is obtained for the choices $\ell(t_{0}=20)\simeq 3.6$ and $\alpha=0.334$. Note that $\ell(t_{0})$ in our analysis is a bare length independent of time and the scaling behavior (5) will be obtained when this is chosen appropriately. These numbers, as expected, provide a constant value of $y(x)$ in the region unaffected by finite system size, which should be identified with the growth amplitude A. The arrow in Fig.\ref{fig3} marks the location where $y(x)$ starts deviating from its constant value, signaling the appearance of finite-size effect at $\ell(t)=(0.75\pm 0.05)\ell_{\max}$. This value is substantially large compared to expectation. Note that the snapshot at $4.5\times10^{6}$ MCS in Fig.\ref{fig1} corresponds to this length.
\par 

Considering the limiting behaviors (\ref{6}) and (\ref{7}), we construct the following functional form of $y(x)$ as

\begin{eqnarray}\label{8}
 y(x)=\frac{Ax}{x+1/(p+qx^{\beta})}.
\end{eqnarray}
The continuous line in Fig.\ref{fig3} is a fit to the form (\ref{8}) with $A\simeq0.294$, $p=3$, $q=6400$, $\beta= 4$, which has the convergence $\lim_{x \rightarrow \infty} y(x)\approx A[1-fx^{-n}]$; $n=5$. Of course, an exponential behavior can also not be ruled out. This behavior may be compared with much slower convergence of such scaling function in dynamic critical phenomena \cite{16}. In the inset of Fig.\ref{fig3}, we plot $[\ell(t)-\ell(t_{0})]^{-3}$ vs $t^{-1}$  for $L=64$ on a log-scale to bring visibility to a wide range of data. The continuous line there is a plot of the form $ax$ with $a\simeq 39=1/A^{3}$. The linear behavior of data starting from very early time justifies the introduction of $\ell(t_{0})$ again. 
\begin{figure}[htb]
%\centering
\includegraphics*[width=0.48\textwidth]{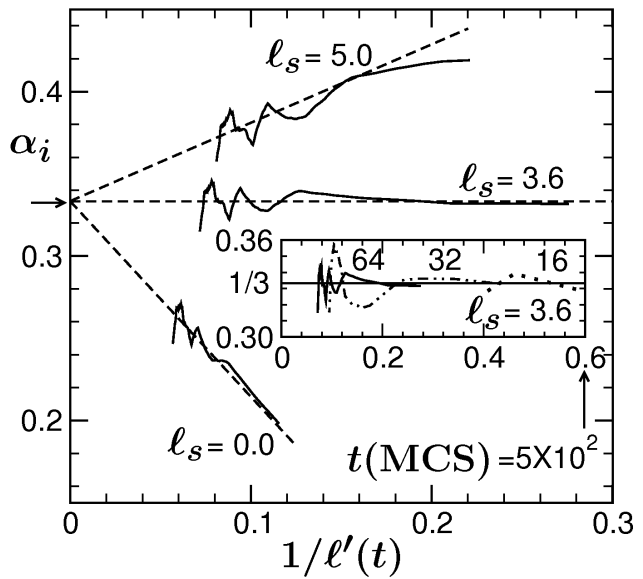}
\caption{\label{fig4} Plot of instantaneous exponent $\alpha_{i}$ as a function of $1/\ell'(t)$ for three different choices of $\ell_s$, for $L=64$. The dashed straight lines have slopes $-1.19, 0$, and $0.49$, respectively. The arrow on the ordinate marks the value $\alpha=1/3$. Inset: $\alpha_{i}$ vs $1/\ell'(t)$ for $\ell_{s}=3.6$ and $L=16,~32$, and $64$.}
\end{figure}
\par
Next we introduce a length $\ell_{s}$ to write
 \begin{eqnarray}\label{9}
  \ell'(t)=\ell(t)-\ell_{s}=[\ell(t_{0})-\ell_{s}]+At^{\alpha},
 \end{eqnarray}
and define an instantaneous exponent $\alpha_{i}=d[\ln \ell'(t)]/d[\ln t]$ to get
\begin{eqnarray}\label{10}
 \alpha_{i}=\alpha\left[1-\frac{\ell(t_{0})-\ell_{s}}{\ell'(t)}\right].
\end{eqnarray}
According to Eq.(\ref{10}), when $\alpha_{i}$ is plotted as a function of $1/\ell'(t)$, asymptotically one expects linear behavior with a $y$-intercept equal to $\alpha$. Fig.\ref{fig4} shows such plots for $\ell_{s}=0.0,~3.6$, and $5.0$, as indicated. The dashed lines have $y$-intercept $\alpha=1/3$ and slopes $m=-[\ell(t_{0})-\ell_{s}]/3$. The consistency of actual data with the dashed lines is remarkable, particularly, the behavior of $\alpha_{i}$ for $\ell_{s}=\ell(t_{0}=20) \simeq 3.6$ again speaks for the choice of $\ell(t_{0})$ and strongly indicates that the LS scaling behavior sets in very early. Also see the inset where we show $\alpha_{i}$ vs $1/\ell'(t)$ for $\ell_{s}\simeq 3.6$ for varying system sizes $L=16,~32$, and $64$. To this end, it is worth mentioning, as suggested by Huse, a term $\propto 1/\ell(t)^{3}$ in Eq.(\ref{2}), could also be motivated by introducing curvature dependence in $\sigma$ as $\sigma/[1+2\delta/\ell(t)]$, $\delta$ being the Tolman length \cite{17}. For a negative value of $\delta$, one then would obtain time-dependent $\alpha_{i}$, as embodied in Eq.(\ref{3}). However, our observation of negligible correction to the exponent, starting from very early time, is consistent with the fact that Tolman length is absent in a symmetrical model \cite{18}. 

\par
The appearance of growing oscillation in $\alpha_{i}$ around the mean value was also pointed out by Shinozaki and Oono \cite{19}. In a finite system, as time increases, for an extended period of time two large neighboring domains of same signs may not merge, thus lowering the value of $\alpha_{i}$, after which when they meet brings in drastic enhancement in $\alpha_{i}$. This character is in fact visible in the direct plot of $\ell(t)$ for $L=128$ in Fig.\ref{fig2}. This oscillation could be a route to an error if one obtains $\alpha$ from least square fitting. However, apart from averaging over huge number of initial configurations, it could also be reduced by considering times well separated from each other while taking instantaneous derivative.
\par
In \textit{conclusion}, this paper contains a comprehensive finite-size scaling analysis of domain coarsening in a phase separating system. Our accurate and appropriate estimate, for which we quote $0.334\pm0.004$, of the growth exponent is almost coincident with the expected LS value $\alpha=1/3$, within tiny error bar. As opposed to the earlier understanding, correction appears to be very small, thus LS scaling regime sets in very early. Very small primary finite-size effect is a welcome message which is suggestive of avoiding large systems, rather focusing on accessing long time scale. Even though size effect may be situation and system dependent, recent study \cite{20} of phase separation in a binary fluid provides qualitative agreement with the finding of the present work. Nevertheless, one should be prepared to encounter stronger size effect in more complicated situations, e.g., systems giving anisotropic patterns \cite{4}.
\par 
In a future paper, we will address the similar issue for phase separation, in different spatial dimensions with varying compositions where larger curvature in domain morphology may delay the convergence to the LS regime. It will be interesting to look at the temperature dependence, in particular, the behavior of the scaling function $y(x)$ as one approaches the critical temperature $T_{c}$ and thus the validity of scaling of $C(r,t)$ for extended period. A comparison of the finite-size effect in non-conserved order parameter situation, e.g., ordering in a ferromagnet, with the present one is also under investigation.
\par
\textit{Acknowledgment}: SKD thanks Professors D. Dhar and S. Puri for useful comments. SM acknowledges Council of Scientific and Industrial Research, India, for financial support in the form of research fellowship.
\par
${*}$ das@jncasr.ac.in

\end{document}